\documentclass[%
 reprint,
nofootinbib,
 amsmath,amssymb,
 aps,
]{revtex4-1}

\usepackage{blindtext}
\usepackage{amsthm}

\usepackage{graphicx}
\usepackage{dcolumn}
\usepackage{bm}

\begin{document}

\title{Nodal variational principle for excited states}

\author{Federico Zahariev}
\email{fzahari@iastate.edu}
\affiliation{Department of Chemistry and Ames Laboratory, Iowa State University,\\
  Ames, Iowa 50011, USA}
\author{Mark S. Gordon}
\email{mark@si.msg.chem.iastate.edu}
\affiliation{Department of Chemistry and Ames Laboratory, Iowa State University,\\ 
  Ames, Iowa 50011, USA}
\author{Mel Levy}
\email{mlevy@tulane.edu}
\affiliation{Department of Chemistry, Duke University,\\
  Durham, North Carolina 27708, USA\\
\\  Department of Physics, North Carolina A\&T State University,\\
  Greensboro, North Carolina 27411, USA\\
\\ Department of Chemistry and Quantum Theory Group, 
  Tulane University,\\ New Orleans, Louisiana 70118, USA\\}

\date{July 31, 2018}

\begin{abstract}
It is proven that the exact excited-state wave function and energy may be obtained by minimizing the energy
expectation value of trial wave functions that are constrained only to have the correct nodes of the state of
interest. This excited-state nodal minimum principle has the advantage that it requires neither minimization
with the constraint of wave-function orthogonality to all lower eigenstates nor the antisymmetry of the trial
wave functions. It is also found that the minimization over the entire space can be partitioned into several
interconnected minimizations within the individual nodal regions, and the exact excited-state energy may be
obtained by a minimization in just one or several of these nodal regions. For the proofs of the theorem, it is
observed that the many-electron eigenfunction (excited state as well as ground state), restricted to a nodal region,
is equivalent to a ground-state wave function of one electron in a higher-dimensional space; and, alternatively, an
explicit excited-state energy variational expression is utilized by generalizing the Jacobi method of multiplicative
variation. In corollaries, error functions are constructed for cases for which the nodes are not necessarily exact. The
exact nodes minimize the energy error functions with respect to nodal variations. Simple numerical illustrations
of the error functions are presented.
\end{abstract}

\maketitle
\section{INTRODUCTION}

Variational principles have provided the most popular and effective ways to compute the properties of electronic systems.
In this connection, it is well known that the minimization of the expectation value of the Hamiltonian yields the wave function
and energy of the $k$th eigenstate, if the trial wave function for the $k$th state is constrained to be orthogonal to the
wave functions for the 0, 1, 2, \ldots{}, $k$-1 states, where the energy of state $n$+1 is understood to be at least as high as the energy of 
state $n$. A related notion is the Hylleraas-Undheim-MacDonald theorem [1]. This theorem states that the eigenvalues of the
Hamiltonian matrix in any finite-dimensional subspace of the Hilbert space are bounded from below by the true eigenvalues
of the Hamiltonian. High-quality results typically require relatively large finite-dimensional subspaces, where the eigenvalue
problem becomes computationally expensive. In fact, the computational cost of the best eigenvalue solver algorithms scales quadratically with the dimension of the subspace.

With this in mind, it is the purpose of this paper to present a nodal variational principle for excited states. Specifically, we
prove that in order to obtain the energy and wave function of the $k$th state, it is sufficient that the minimization takes place with
the constraint that the trial wave function has the same nodes as the wave function of the $k$th eigenstate. It is not necessary
to impose the difficult orthogonality constraint. It is also not necessary to impose explicitly antisymmetry. The imposition
of the nodal constraint is sufficient.

While interest in nodes of eigenfunctions goes back at least
to the proof that the $k$th eigenfunction of the one-electron
Schrödinger equation, in any multidimensional space, has no
more than $k$ nodal regions [2], and although research regarding
nodes and their properties continued [3], it is the ground state
fixed-node variational principle [4] and tiling theorem
[5] of the Quantum Monte Carlo (QMC) method that has
aroused substantial interest in nodes and their properties [6?11]. The ground-state fixed-node variational principle states
that an energy minimization in a nodal region of an arbitrary
antisymmetric wave function gives an upper bound to the
ground-state energy, and if a nodal region is bounded by the
exact nodes, the energy minimization gives the ground-state
energy. The proof of the ground-state fixed-node variational
principle indirectly relies on the tiling theorem [5].

The QMC method is now being commonly used for excited states as well as ground states. In fact, the nodal variational
principle for excited states presented in this paper is being implied \emph{without a proof} for a number of QMC applications,
such as the computations of optical gaps in nanostructures [12] and solids [13], diffusive properties of the vacancy defects in
diamond [14], diamonoid excitation energies and Stokes shifts [15], excitation spectra of localized Wigner states [16], quasiparticle
excitations of the electron gas [17], and electronic [18] and rovibrational excitations [19] of molecules. As the QMC
experience demonstrates, even approximations to the correct nodal surfaces typically result in accurate excited-state values.
 
 The ground-state fixed-node variational principle has been extended to nondegenerate [6] and degenerate [7] excited
states that are ground states within certain symmetry classes of trial wave functions. More precisely, the trial wave functions
are supposed to transform according to the one-dimensional irreducible representation of the symmetry point group of the
Hamiltonian. The proofs that are used therein are symmetry restricted generalizations of the ground-state fixed-node proof
[4] and rely on symmetry-restricted generalizations of the ground-state tiling theorem [5]. Although symmetries are not
uncommon in molecules consisting of a handful of atoms, larger molecules are less likely to possess any symmetry, and
no tiling theorem currently exists that would be applicable to an arbitrary excited state. The proofs of the theorem, its
corollaries, and the supporting lemma in the current paper do not require a tiling theorem and are applicable to any
eigenstate. In contrast with the original fixed-node approach though, variance type error functions are minimized here for
the case of approximate nodes.

We prove the theorem and its corollaries by means of two complementary routes, A and B. Proof A is based on
our observation that a many-electron wave function, with a domain of definition that is restricted to a single nodal
region, is equivalent to a single-electron wave function in a higher-dimensional space. Proof B extends the ground-state
Jacobi method of multiplicative variation to excited states.

Moreover, when the exact nodes are not known, corollaries to the proofs given here construct two different error functions that assess the quality of approximate nodes. These error functions incorporate energy minimization with the given
approximate nodes. The minimization of the error functions, with respect to variations of the nodes, achieves zero once
the geometries of the nodes become exact. We show that \emph{the explicit antisymmetry is not required}, even when the nodes are
approximate. In fact, it is important to emphasize that we prove in a lemma that the minimization results in an antisymmetric
wave function. Numerical examples illustrate the use of the error functions.\\

\section{NODAL VARIATIONAL PRINCIPLE}
 
Given below are two different proofs of our \emph{theorem} that
expresses the following nodal variational principle for excited states:

(i) \emph{The minimum of the energy expectation value of trial wave functions
that are analytically well behaved and have the nodes of the exact eigenfunction}
$\Psi_k(\textbf r_1,\textbf r_2,...,\textbf r_N)$ \emph{ of
N-electrons is the exact eigenvalue}
$E_k$\emph{. The
minimum of the energy expectation value is achieved at the exact
eigenfunction}
$\Psi_k(\textbf r_1,\textbf r_2,...,\textbf r_N)$\emph{.}

(ii) \emph{In
addition, even the minimization in just one or several nodal regions also
yields} $E_k$\emph{. }

Note that it has been shown [20, 21] that spin-free wave functions are sufficient in the context of the present work.

\section{PROOFS OF THE THEOREM}

\begin{proof}[Proof A] Consider the nodal hypersurface corresponding to the $k$th eigenfunction,
i.e., all of the points in the 3$N$-dimensional coordinate space of $N$
electrons that satisfy the condition $\Psi_k(\textbf r_1,\textbf r_2,...,\textbf r_N)=0$. 
This nodal hypersurface, i.e. a (3$N$-1)-dimensional surface in the
3$N$-dimensional space of electron positions, partitions the configuration
space into $m$ nodal regions $L_j\ (j=1,2,...,m)$. $\Psi_k(\textbf r_1,\textbf r_2,...,\textbf r_N)$ 
is either strictly positive or strictly negative in each of the $m$ nodal regions. Some technical aspects of the nodal constraint are in Appendix A.

Now consider a trial wave function $\Psi^{(k)}(\textbf r_1,\textbf r_2,...,\textbf r_N)$  that is not necessarily antisymmetric with respect to the exchange of like-spin electrons and has the same nodes as the $k$th eigenfunction $\Psi_k(\textbf r_1,\textbf r_2,...,\textbf r_N)$. The trial
wave function $\Psi^{(k)}(\textbf r_1,\textbf r_2,...,\textbf r_N)$, which is normalized to unity, could be the exact $k$th eigenfunction 
$\Psi_k(\textbf r_1,\textbf r_2,...,\textbf r_N)$  itself. The
integration over the entire 3$N$-dimensional space can be partitioned into
a sum of integrations over the $m$ nodal regions,

\begin{equation}\label{1}
\langle\Psi^{(k)}|\Psi^{(k)}\rangle=\sum_{j=1}^m\langle\Psi^{(k)}|\Psi^{(k)}\rangle_{L_j}=\sum_{j=1}^mp_{L_j}=1,
\end{equation}

\noindent where $\langle\Psi^{(k)}|\Psi^{(k)}\rangle_{L_j}$ signifies
$\langle\Psi^{(k)}|\Psi^{(k)}\rangle$ in the nodal
region $L_j$. 

The energy expectation value of $\Psi^{(k)}(\textbf r_1,\textbf r_2,...,\textbf r_N)$ can be similarly partitioned as

\begin{widetext}
\begin{equation}\label{2}
E^{(k)}=\langle\Psi^{(k)}|\hat H|\Psi^{(k)}\rangle=
\sum_{j=1}^m\langle\Psi^{(k)}|\Psi^{(k)}\rangle_{L_j}\frac{\langle\Psi^{(k)}|\hat H|\Psi^{(k)}\rangle_{L_j}}{\langle\Psi^{(k)}|\Psi^{(k)}\rangle_{L_j}}.
\end{equation}
\end{widetext}

The expression $\frac{\langle\Psi^{(k)}|\hat H|\Psi^{(k)}\rangle_{L_j}}{\langle\Psi^{(k)}|\Psi^{(k)}\rangle_{L_j}}$, which we denote as $E_{L_j}^{(k)}$, on
the right-hand side of Eq. (\ref{2}) is the energy expectation value of
$\Psi^{(k)}(\textbf r_1,\textbf r_2,...,\textbf r_N)$ in the
individual nodal regions
$L_j$ and $p_{L_j}=\langle\Psi^{(k)}|\Psi^{(k)}\rangle_{L_j}$
is the respective probability of finding the $N$-electron system in the
individual nodal region
$L_j$. Consequently,
the right-hand side of Eq. (\ref{2}) is an average over the nodal energies
that are weighted by the respective probabilities. If the trial
wave function $\Psi^{(k)}(\textbf r_1,\textbf r_2,...,\textbf r_N)$
is the exact eigenfunction
$\Psi_k(\textbf r_1,\textbf r_2,...,\textbf r_N)$ itself, then
$E_{L_j}^{(k)}=E^{(k)}=E_k$. (A similar
partitioning of the energy expectation value of a one-dimensional
Hamiltonian was used in Ref. [11] in the proof of a different
variational principle involving nodes.)

It is important to observe here that the $k$th eigenfunction
$\Psi_k(\textbf r_1,\textbf r_2,...,\textbf r_N)$ in a nodal
region is, in fact, the ground-state solution for the given nodal
region. This is because an eigenfunction that is either strictly
positive or strictly negative is a ground state according to an
extension presented here of a theorem of Courant and Hilbert {[}2{]}. 
Although the original theorem is for a one-electron wave function in a
space of arbitrary dimension, \emph{the many-electron eigenfunction
$\Psi_k(\textbf r_1,\textbf r_2,...,\textbf r_N)$, restricted to
a nodal region, can be equivalently interpreted as a ground state
wave function of one electron in 3$N$-dimensional space, even when $\Psi_k(\textbf r_1,\textbf r_2,...,\textbf r_N)$, 
is an excited state}. \footnote{Note that  the interchange symmetry of
  $\Psi_k(\textbf r_1,\textbf r_2,...,\textbf r_N)$
  does not play a role for an individual nodal region for the following
  reason. If
  $\textbf r_1,\textbf r_2 ,...,\textbf r_i ,...,\textbf r_j,...,\textbf r_N $ 
  belongs to a nodal region, then
  $\textbf r_1,\textbf r_2,..., \textbf r_j,..., \textbf r_i,...,\textbf r_N$,
  in which the spatial coordinates corresponding to two spin-equivalent
  electrons are interchanged, is outside the nodal region, as the
  interchange changes the sign of the wave function.}
  
In such an interpretation, the many-electron Hamiltonian is regarded as an
effective Hamiltonian of one electron in 3$N$-dimensional space and, 
similarly, $\Psi_k(\textbf r_1,\textbf r_2,...,\textbf r_N)$
may also be regarded as an eigenfunction of one electron in
3$N$-dimensional space.

According to the foregoing ground state minimum principle for each nodal
region, the nodal region normalized energy expectation value of $\Psi^{(k)}(\textbf r_1,\textbf r_2,...,\textbf r_N)$ cannot be
lower than the nodal region normalized energy expectation value of the
$k$th eigenvalue
of $\Psi_k(\textbf r_1,\textbf r_2,...,\textbf r_N)$:

\begin{equation}\label{3}
E_{L_j}^{(k)}=\frac{\langle\Psi^{(k)}|\hat H|\Psi^{(k)}\rangle_{L_j}}{\langle\Psi^{(k)}|\Psi^{(k)}\rangle_{L_j}}
\ge \frac{\langle\Psi_k|\hat H|\Psi_k\rangle_{L_j}}{\langle\Psi_k |\Psi_k\rangle_{L_j}}=E_k.
\end{equation}

Multiplication on both sides of the inequality in Eq. (\ref{3}) by
$p_{L_j}$ followed by a
summation over $j$ gives

\begin{equation}\label{4}
\begin{split}
E^{(k)}&=\langle\Psi^{(k)}|\hat H|\Psi^{(k)}\rangle=\sum_{j=1}^mp_{L_j}E_{L_j}^{(k)}\ge\sum_{j=1}^mp_{L_j}E_k\\
&=(\sum_{j=1}^mp_{L_j})E_k=E_k.
\end{split}
\end{equation}

\noindent The inequality in Eq. (\ref{4}) arises because each
$p_{L_j}$ is
non-negative, the use of the normalization expression given by Eq. (\ref{1}), and the fact that the weighted average increases if any of the contributing energies increases. Equation (\ref{4}) proves part (i) of the theorem. 

The analytic restriction on the trial wave functions guarantees \emph{smooth patching} of the wave functions that achieve energy 
minima in the different nodal regions. This smooth patching is \emph{necessary} because each nodal-region energy minimizing wave 
function has the freedom of a multiplicative constant.

Equation (\ref{3}) demonstrates that an energy minimization in an isolated nodal
region actually gives the exact energy $E_k$ of the entire
eigenfunction $\Psi_k(\textbf r_1,\textbf r_2,...,\textbf r_N)$.
More generally, consider an energy minimization over some of the nodal
regions, such as over an isolated region of space bounded by nodes. An
appropriately normalized nodal energy minimization over just some of the
nodal regions also yields the exact energy $E_k$, as
demonstrated by a generalization of Eq. (\ref{4}),

\begin{equation}\label{5}
\frac{\sum_jp_{L_j}E_{L_j}^{(k)}}{\sum_jp_{L_j}}\ge\frac{\sum_jp_{L_j}E_k}{\sum_jp_{L_j}}=E_k,
\end{equation}

\noindent where the partial sum is only over one or more nodes that participate in the
minimization. Eq. (\ref{5}) proves part (ii) of the above theorem.
\end{proof}

\begin{proof}[Proof B] Consider trial wave functions of the type
$g(\textbf r_1,\textbf r_2,...,\textbf r_N)\Psi_k(\textbf r_1,\textbf r_2,...,\textbf r_N)$, where the
$k$th state
$\Psi_k(\textbf r_1,\textbf r_2,...,\textbf r_N)$ is kept fixed
and the function
$g(\textbf r_1,\textbf r_2,...,\textbf r_N)$ is varied. The
function $g(\textbf r_1,\textbf r_2,...,\textbf r_N)$ is
assumed to be well-behaved. That is, $g(\textbf r_1,\textbf r_2,...,\textbf r_N)$  is smooth (in particular, everywhere finite) 
and such that
$g(\textbf r_1,\textbf r_2,...,\textbf r_N)\Psi_k(\textbf r_1,\textbf r_2,...,\textbf r_N)$ is a
well-behaved wave function. It is important to note that
$g(\textbf r_1,\textbf r_2,...,\textbf r_N)\Psi_k(\textbf r_1,\textbf r_2,...,\textbf r_N)$ is not assumed
here to be necessarily antisymmetric with respect to the exchange of
like-spin electrons.

The theorem will now be proven by showing that the explicit form of the
$g$-variations around the excited state
$\Psi_k(\textbf r_1,\textbf r_2,...,\textbf r_N)$, which can be
considered to be a generalization to excited states of the Jacobi method
of multiplicative variation,\footnote{On pp. 458 and 459 of Vol. I of Ref [2], the Jacobi's 
method of multiplicative variation is introduced and applied to the ground-state 
problem only.} is

\begin{widetext}
\begin{equation}\label{6}
\frac{\langle g\Psi_k|\hat H|g\Psi_k\rangle}{\langle g\Psi_k|g\Psi_k\rangle}=E_k+\frac{\frac{1}{2}\sum_{i=1}^N\sum_{\alpha=x,y,z}\langle (\frac{\partial g}{\partial r_{i,\alpha}})\Psi_k|\hat H|(\frac{\partial g}{\partial r_{i,\alpha}})\Psi_k\rangle}{\langle g\Psi_k|g\Psi_k\rangle}\ge E_k.
\end{equation}
\end{widetext}

\noindent Note that the inequality in Eq. (\ref{6}) occurs because the sums are
non-negative.

The equality on the left in Eq. (\ref{6}) is derived by the following chain of
equalities

\begin{widetext}
\begin{align}\label{7}
\frac{\langle g\Psi_k|\hat H|g\Psi_k\rangle}{\langle g\Psi_k|g\Psi_k\rangle}
&=\frac{\langle g\Psi_k|\hat T|g\Psi_k\rangle+\langle g\Psi_k|\hat V|g\Psi_k\rangle}{\langle g\Psi_k|g\Psi_k\rangle}
=\frac{\langle g\Psi_k|\hat T|g\Psi_k\rangle+\langle g^2\Psi_k|\hat V|\Psi_k\rangle}{\langle g\Psi_k|g\Psi_k\rangle}\\
&=\frac{\langle g\Psi_k|\hat T|g\Psi_k\rangle+\langle g^2\Psi_k|(\hat H-\hat T)|\Psi_k\rangle}{\langle g\Psi_k|g\Psi_k\rangle}
=E_k+\frac{\langle g\Psi_k|\hat T|g\Psi_k\rangle-\langle g^2\Psi_k|\hat T|\Psi_k\rangle}{\langle g\Psi_k|g\Psi_k\rangle}\nonumber\\
&=E_k+\frac{\frac{1}{2}\sum_{i=1}^N\sum_{\alpha=x,y,z}\langle (\frac{\partial g}{\partial r_{i,\alpha}})\Psi_k|\hat H|
(\frac{\partial g}{\partial r_{i,\alpha}})\Psi_k\rangle}{\langle g\Psi_k|g\Psi_k\rangle}\nonumber.
\end{align}
\end{widetext}

\noindent Additional details of the derivation of Eq. (\ref{7}) can be found in Appendix B.

At this stage, the inequality in Eq. (\ref{6}) has been proved. But in order
for the inequality to constitute a proof of the theorem, each trial
wave function $\Psi^{(k)}(\textbf r_1,\textbf r_2,...,\textbf r_N)$,
that has the same nodes as the $k$th eigenfunction
$\Psi_k(\textbf r_1,\textbf r_2,...,\textbf r_N)$, should be
presentable as
$g(\textbf r_1,\textbf r_2,...,\textbf r_N)\Psi_k(\textbf r_1,\textbf r_2,...,\textbf r_N)$. In other
words, the well-behaved scaling function must be presentable as $\frac{\Psi^{(k)}(\textbf r_1,\textbf r_2,...,\textbf r_N)}
{\Psi_k(\textbf r_1,\textbf r_2,...,\textbf r_N)}$. Since
$\Psi_k(\textbf r_1,\textbf r_2,...,\textbf r_N)$ vanishes at
the nodes, the finiteness of the ratio may not appear to be guaranteed.
However, the ratio is, in fact, finite, as shown in Appendix C.

Thus, the inequality in Eq. (\ref{6}), together with the fact that each trial wave 
function $\Psi^{(k)}(\textbf r_1,\textbf r_2,...,\textbf r_N)$ 
that has the same nodes as the $k$th eigenfunction
$\Psi_k(\textbf r_1,\textbf r_2,...,\textbf r_N)$ is presentable
as $g(\textbf r_1,\textbf r_2,...,\textbf r_N)\Psi_k(\textbf r_1,\textbf r_2,...,\textbf r_N)$, proves part (i) 
of the above theorem.

As with Proof A, Proof B can be adapted to a single nodal region or,
more generally, to several nodal regions with an appropriate
normalization of the energy expectation value. Equation (\ref{6}) implies that
the analog of Eq. (\ref{5}) is

\begin{widetext}
\begin{equation}\label{8}
\frac{\sum_j\langle g\Psi_k|\hat H|g\Psi_k\rangle_{L_j}}{\sum_j\langle g\Psi_k|g\Psi_k\rangle_{L_j}}
=E_k+\frac{\frac{1}{2}\sum_j\sum_{i=1}^N\sum_{\alpha=x,y,z}\langle (\frac{\partial g}{\partial r_{i,\alpha}})
\Psi_k|\hat H|(\frac{\partial g}{\partial r_{i,\alpha}})\Psi_k\rangle_{L_j}}{\sum_j\langle g\Psi_k|g\Psi_k\rangle_{L_j}}\ge E_k,
\end{equation}
\end{widetext}

\noindent where each sum in $j$ could be replaced be simply one term when only one nodal region is used,
which proves part (ii) of the above theorem.

\end{proof}

\section{ANTISYMMETRIC LEMMA AND COROLLARIES TO THE THEOREM}

Now assume that the $m$ nodal regions $\tilde{L}_j\ (j=1,2,...,m)$ are not necessarily the exact nodes 
of $\Psi_k$. It is assumed that the approximate nodes are variations around the exact ones, i.e., that 
the approximate nodes can be continuously deformed back to the exact ones.  

In this case, the minimizing energies within the different nodal regions,

\begin{equation}\label{9}
\tilde{E}_{\tilde{L}_j,min}^{(k)}=\frac{\langle\tilde{\Psi}_{min}^{(k)}|\hat H|\tilde{\Psi}_{min}^{(k)}\rangle_{\tilde{L}_j}}{\langle\tilde{\Psi}_{min}^{(k)}|\tilde{\Psi}_{min}^{(k)}\rangle_{\tilde{L}_j}},
\end{equation}

\noindent may differ from each other. Although the trial wave functions $\tilde{\Psi}^{(k)}$ are not \emph{constrained} to be antisymmetric, 
the energy-minimizing trial wave function $\tilde{\Psi}_{min}^{(k)}$ will always be antisymmetric if the nodes come from \emph{some} 
antisymmetric wave function, as the following lemma demonstrate. 

\emph{Lemma.} The minimizing wave function $\tilde{\Psi}_{min}^{(k)}$ is antisymmetric.
(The spin-free wave functions that are antisymmetric are such with respect to the interchange of electron coordinates that 
correspond to the same spin.)

\begin{proof} Define $\Phi$ to be the antisymmetric wave function such that the nodes of $\Phi$ 
divide the $N$-electron configuration space into $m$ nodal regions $\tilde{L}_j$ ($j=1,2,...,m$). The nodes of the trial wave functions $\tilde{\Psi}^{(k)}$ 
are assumed to be the nodes of $\Phi$.

Choose a point $\vec{R}=(\vec{r}_1,\vec{r}_2,...,\vec{r}_N)$ in the
configuration space of $N$ electrons. Label the nodal region, where the point $\vec{R}$
lies, as $A$. An interchange of two electrons having the same spin, say the first and 
the second electrons, maps the point $\vec{R}$ to a new point $\vec{R}'=(\vec{r}_2,\vec{r}_1,...,\vec{r}_N)$.
Label the nodal region, where the point $\vec{R}'$ lies, as $A'$.

The nodal regions $A$ and $A'$ are different, because $\tilde{\Phi}(\vec{r}_1,\vec{r}_2,...,\vec{r}_N)$ 
and $\tilde{\Phi}(\vec{r}_2,\vec{r}_1,...,\vec{r}_N)$ have different signs (as a reminder: $\Phi$
is antisymmetric). If $\vec{R}$ and $\vec{R}'$ are connected with a straight 
line, there has to be an odd number of nodal crossings along the line as there is a sign change 
at each nodal crossing.

The interchange of the first and second electrons, in fact, maps every point of $A$ to a 
point of $A'$ making the two nodal regions "isomorphic", i.e. of the same form and size. Since 
the nodal regions $A$ and $A'$ are isomorphic, the ground state in $A$ is mapped to the ground 
state in $A'$ by the interchange of the first and second electrons (up to a normalization 
factor). In the same manner, another nodal region, say $B$, is mapped to an isomorphic 
nodal region $B'$, $C$ to $C'$, $D$ to $D'$ and so on. In other words, one half of the space 
$(A, B, C, D, ...)$ is mapped to its isomorphic other half $(A', B', C', D', ...)$. The uncertainty in the 
normalization factor of the ground state is reduced to just an uncertainty in the sign due to the 
perfect mirror symmetry between the two isomorphic halves.

The minimizing wave function $\tilde{\Psi}_{min}^{(k)}$ is a ground state within 
each nodal region. As a result, $\tilde{\Psi}_{min}^{(k)}$ restricted to $A$ is mapped 
to $\tilde{\Psi}_{min}^{(k)}$ restricted to $A'$ by the interchange of the first and
second electrons, i.e. $\tilde{\Psi}_k(\vec{r}_1,\vec{r}_2,...,\vec{r}_N)
=\pm\tilde{\Psi}_k(\vec{r}_2,\vec{r}_1,...,\vec{r}_N)$.

On the one hand, every minimizing wave function, antisymmetric or not, changes sign 
across a node because it is with a nonzero slope at the node (Appendix C). On the other hand, as 
stated above, there is an odd number of nodal crossings along the straight line 
connecting $\vec{R}$ and $\vec{R}'$. Hence, $\tilde{\Psi}_k(\vec{r}_1,\vec{r}_2,...,\vec{r}_N)
=-\tilde{\Psi}_k(\vec{r}_2,\vec{r}_1,...,\vec{r}_N)$.

\end{proof}

A relevant ``error expression'', corresponding to $\tilde{\Psi}_{min}^{(k)}$, is

\begin{equation}\label{10}
\sum_{j=1}^m\big[\tilde{E}_{\tilde{L}_j,min}^{(k)}-\tilde{E}_{min}^{(k)}\big]\langle\tilde{\Psi}_{min}^{(k)}|\tilde{\Psi}_{min}^{(k)}\rangle_{\tilde{L}_j},
\end{equation}

\noindent where

\begin{align}\label{11}
\tilde{E}_{min}^{(k)}&=\langle\tilde{\Psi}_{min}^{(k)}|\hat H|\tilde{\Psi}_{min}^{(k)}\rangle\nonumber\\
&=\sum_{j=1}^m\tilde{E}_{\tilde{L}_j,min}^{(k)}\langle\tilde{\Psi}_{min}^{(k)}|\tilde{\Psi}_{min}^{(k)}\rangle_{\tilde{L}_j}.
\end{align}

Note that the larger nodal regions are weighted higher in expression (\ref{10}). This error expression achieves its minimum of zero 
if and only if the trial wave function $\tilde{\Psi}_{min}^{(k)}$ is the true eigenfunction $\Psi_k$, because then all the nodal-region minimizing 
energies in Eq. (\ref{9}) are equal.

\emph{Corollary I to the theorem.}  The minimization of error expression (\ref{10}), with respect to nodal variations, yields the correct nodes 
of $\Psi_k$. 

Another nodal error expression is dictated by the use of $g\Psi_k$ in Proof B. The expression is 

\begin{equation}\label{12}
\Bigg[\frac{\langle g\tilde{\Psi}_{min}^{(k)}|\hat H|\tilde{\Psi}_{min}^{(k)}\rangle}{\langle g\tilde{\Psi}_{min}^{(k)}|\tilde{\Psi}_{min}^{(k)}\rangle}
-\langle\tilde{\Psi}_{min}^{(k)}|\hat H|\tilde{\Psi}_{min}^{(k)}\rangle\Bigg]^2.
\end{equation}

\emph{Corollary II to the Theorem.}  Minimization of error expression (\ref{12}) with respect to nodal variations, 
for all allowable scaling functions $g$, yields the correct nodes of $\Psi_k$. Note that the allowable $g$'s from 
Proof B are such that $g\tilde{\Psi}_{min}^{(k)}$ preserves the nodes of $\tilde{\Psi}_{min}^{(k)}$.

When only a subset of the nodal regions is employed, for which zero values of error expressions (10)-(12) serve as 
necessary eigenstate conditions, it is understood that expressions (10)-(12) are adjusted to incorporate the particular scalar products 
in the nodal regions.

\section{SIMPLE NUMERICAL EXAMPLES}

\begin{table*}
\caption{The energies and corresponding errors of the wave function for the $4S$ state of the hydrogen atom with approximate nodes. 
The energies are in Hartrees. The squared norms of the wave function in the four nodal regions are, respectively, 0.007 188, 0.030 936, 0.128 878, and
0.832 998.}
\begin{ruledtabular}
\begin{tabular}{ccccc}
Nodal & $\tilde{E}_{min}^{(k)}$ & $[E_k-\tilde{E}_{min}^{(k)}]^2$& Eq. (\ref{10}) with   & Eq. (\ref{10}) with \\
regions&&with $\tilde{E}_{min}^{(k)}$  over &$\tilde{E}_{min}^{(k)}$ over&$\tilde{E}_{min}^{(k)}$ over the\\
&&the nodal regions&1,2,3,4&nodal regions \\
\hline 
1 &	-0.14010	& 1.1849$\times 10^{-2}$ & 1.1819$\times 10^{-2}$ & 0  \\
2 &	-0.01000	& 4.5153$\times 10^{-4}$ & 4.5743$\times 10^{-4}$ & 0  \\
3 &	-0.03261	& 1.8427$\times 10^{-6}$ & 1.4862$\times 10^{-6}$ & 0  \\
4 &	-0.03106	& 3.7671$\times 10^{-8}$ & 1.1052$\times 10^{-7}$ & 0  \\
1,2 &	 -0.03453	& 1.0764$\times 10^{-5}$ & 2.5995$\times 10^{-3}$ & 2.5897$\times 10^{-3}$ \\
1,3 & -0.03829	& 4.9508$\times 10^{-5}$ & 6.2577$\times 10^{-4}$ & 5.7819$\times 10^{-4}$ \\
1,4 &	 -0.03199	& 5.4590$\times 10^{-7}$ & 1.0122$\times 10^{-4}$ & 1.0086$\times 10^{-4}$ \\
2,3 & -0.02823	& 9.1122$\times 10^{-6}$ & 8.9746$\times 10^{-5}$ & 7.9779$\times 10^{-5}$\\
2,4 & -0.03030  & 8.9878$\times 10^{-7}$ & 1.6486$\times 10^{-5}$ & 1.5306$\times 10^{-5}$ \\
3,4 & -0.03126  & 1.9032$\times 10^{-10}$ & 2.9484$\times 10^{-7}$ & 2.7933$\times 10^{-7}$ \\
1,2,3	 & -0.03305 & 3.2276$\times 10^{-6}$ & 5.9458$\times 10^{-4}$	& 5.9184$\times 10^{-4}$ \\
1,2,4	 & -0.03121 & 1.7651$\times 10^{-9}$ & 1.1387$\times 10^{-4}$	& 1.1384$\times 10^{-4}$ \\
1,3,4	 & -0.03207 & 6.7424$\times 10^{-7}$ & 8.7959$\times 10^{-5}$	& 8.7493$\times 10^{-5}$ \\
2,3,4	 & -0.03060 & 4.2089$\times 10^{-7}$ & 1.4539$\times 10^{-5}$	& 1.3920$\times 10^{-5}$ \\
1,2,3,4 & -0.03139 &	1.9140$\times 10^{-8}$ & 1.0167$\times 10^{-4}$ & 1.0167$\times 10^{-4}$ \\
\end{tabular}
\end{ruledtabular}
\end{table*}

\begin{table*}
\caption{The energies and corresponding errors of the first alternative wave function for the $4S$ state of the hydrogen atom with approximate nodes. 
The energies are in Hartrees. The squared norms of the wave function in the four nodal regions are, respectively, 0.003 377, 0.029 859, 0.096 241, and 0.870 523.}
\begin{ruledtabular}
\begin{tabular}{ccccc}
Nodal & $\tilde{E}_{min}^{(k)}$ & $[E_k-\tilde{E}_{min}^{(k)}]^2$& Eq. (\ref{10}) with   & Eq. (\ref{10}) with \\
regions&&with $\tilde{E}_{min}^{(k)}$  over &$\tilde{E}_{min}^{(k)}$ over&$\tilde{E}_{min}^{(k)}$ over the\\
&&the nodal regions&1,2,3,4&nodal regions \\
\hline 
1 &	-0.01000	& 4.5153$\times 10^{-4}$ & 4.5442$\times 10^{-4}$ & 0  \\
2 &	-0.04337	& 1.4700$\times 10^{-4}$ & 1.4536$\times 10^{-4}$ & 0  \\
3 &	-0.02636	& 2.3944$\times 10^{-5}$ & 2.4611$\times 10^{-5}$ & 0  \\
4 &	-0.03153	& 8.1424$\times 10^{-8}$ & 4.7355$\times 10^{-8}$ & 0  \\
1,2 &	 -0.03998	& 7.6272$\times 10^{-5}$ & 1.7676$\times 10^{-4}$ & 1.0167$\times 10^{-4}$ \\
1,3 & -0.02580	& 2.9677$\times 10^{-5}$ & 3.9181$\times 10^{-5}$ & 8.7611$\times 10^{-6}$ \\
1,4 &	 -0.03145	& 4.0860$\times 10^{-8}$ & 1.8031$\times 10^{-6}$ & 1.7850$\times 10^{-6}$ \\
2,3 & -0.03039	& 7.5963$\times 10^{-7}$ & 5.3203$\times 10^{-5}$ & 5.2335$\times 10^{-5}$\\
2,4 & -0.03193  & 4.5963$\times 10^{-7}$ & 4.8663$\times 10^{-6}$ & 4.4939$\times 10^{-6}$ \\
3,4 & -0.03102  & 5.2981$\times 10^{-8}$ & 2.4927$\times 10^{-6}$ & 2.4039$\times 10^{-6}$ \\
1,2,3	 & -0.02985 & 1.9470$\times 10^{-6}$ & 6.3667$\times 10^{-5}$	& 6.1526$\times 10^{-5}$ \\
1,2,4	 & -0.03185 & 3.5525$\times 10^{-7}$ & 6.5460$\times 10^{-6}$	& 6.2670$\times 10^{-6}$ \\
1,3,4	 & -0.03095 & 9.2014$\times 10^{-8}$ & 4.0657$\times 10^{-6}$	& 3.9280$\times 10^{-6}$ \\
2,3,4	 & -0.03139 & 1.9591$\times 10^{-8}$ & 6.7730$\times 10^{-6}$	& 6.7678$\times 10^{-6}$ \\
1,2,3,4 & -0.03138 &	4.5884$\times 10^{-9}$ & 9.4020$\times 10^{-6}$ & 9.4020$\times 10^{-6}$ \\
\end{tabular}
\end{ruledtabular}
\end{table*}

\begin{table*}
\caption{The energies and corresponding errors of the second alternative wave function for the $4S$ state of the hydrogen atom with approximate nodes. The energies are in Hartrees. The squared norms of the wave function in the four nodal regions are, respectively, 0.005 799, 0.037 427, 0.109 387, and 0.847 386.}
\begin{ruledtabular}
\begin{tabular}{ccccc}
Nodal & $\tilde{E}_{min}^{(k)}$ & $[E_k-\tilde{E}_{min}^{(k)}]^2$& Eq. (\ref{10}) with   & Eq. (\ref{10}) with \\
regions&&with $\tilde{E}_{min}^{(k)}$  over &$\tilde{E}_{min}^{(k)}$ over&$\tilde{E}_{min}^{(k)}$ over the\\
&&the nodal regions&1,2,3,4&nodal regions \\
\hline 
1 &	-0.02075	& 1.1023$\times 10^{-4}$ & 1.1123$\times 10^{-4}$ & 0  \\
2 &	-0.03789	& 4.4137$\times 10^{-5}$ & 4.3511$\times 10^{-5}$ & 0  \\
3 &	-0.02588	& 2.8840$\times 10^{-5}$ & 2.9350$\times 10^{-5}$ & 0  \\
4 &	-0.03178	& 2.7825$\times 10^{-7}$ & 2.3057$\times 10^{-7}$ & 0  \\
1,2 &	 -0.03559	& 1.8868$\times 10^{-5}$ & 5.2595$\times 10^{-5}$ & 3.4136$\times 10^{-5}$ \\
1,3 & -0.02562	& 3.1680$\times 10^{-5}$ & 3.3473$\times 10^{-5}$ & 1.2577$\times 10^{-6}$ \\
1,4 &	 -0.03170	& 2.0479$\times 10^{-7}$ & 9.8504$\times 10^{-7}$ & 8.2084$\times 10^{-7}$ \\
2,3 & -0.02894	& 5.3250$\times 10^{-6}$ & 3.2960$\times 10^{-5}$ & 2.7415$\times 10^{-5}$\\
2,4 & -0.03204  & 6.1811$\times 10^{-7}$ & 2.0613$\times 10^{-6}$ & 1.5154$\times 10^{-6}$ \\
3,4 & -0.03110  & 2.1550$\times 10^{-8}$ & 3.5598$\times 10^{-6}$ & 3.5221$\times 10^{-6}$ \\
1,2,3	 & -0.02863 & 6.8585$\times 10^{-6}$ & 3.5934$\times 10^{-5}$	& 2.8826$\times 10^{-5}$ \\
1,2,4	 & -0.03196 & 5.0796$\times 10^{-7}$ & 2.7722$\times 10^{-6}$	& 2.3294$\times 10^{-6}$ \\
1,3,4	 & -0.03104 & 4.3752$\times 10^{-8}$ & 4.2085$\times 10^{-6}$	& 4.1427$\times 10^{-6}$ \\
2,3,4	 & -0.03136 & 1.1844$\times 10^{-8}$ & 5.0638$\times 10^{-6}$	& 5.0600$\times 10^{-6}$ \\
1,2,3,4 & -0.03130 &	2.2384$\times 10^{-9}$ & 7.2620$\times 10^{-6}$ & 7.2620$\times 10^{-6}$ \\
\end{tabular}
\end{ruledtabular}
\end{table*}

\begin{table*}
\caption{The energies and corresponding errors of the third alternative wave function for the $4S$ state of the hydrogen atom with approximate nodes. 
The energies are in Hartrees. The squared norms of the wave function in the four nodal regions are, respectively, 0.007 466, 0.102 640, 0.240 274, and 0.649 620.}
\begin{ruledtabular}
\begin{tabular}{ccccc}
Nodal & $\tilde{E}_{min}^{(k)}$ & $[E_k-\tilde{E}_{min}^{(k)}]^2$& Eq. (\ref{10}) with   & Eq. (\ref{10}) with \\
regions&&with $\tilde{E}_{min}^{(k)}$  over &$\tilde{E}_{min}^{(k)}$ over&$\tilde{E}_{min}^{(k)}$ over the\\
&&the nodal regions&1,2,3,4&nodal regions \\
\hline 
1 &	-0.01000	& 4.5153$\times 10^{-4}$ & 5.7794$\times 10^{-4}$ & 0  \\
2 &	-0.05227	& 4.4195$\times 10^{-4}$ & 3.3239$\times 10^{-4}$ & 0  \\
3 &	-0.03312	& 3.4957$\times 10^{-6}$ & 8.4886$\times 10^{-7}$ & 0  \\
4 &	-0.03178	& 2.7835$\times 10^{-7}$ & 5.1235$\times 10^{-6}$ & 0  \\
1,2 &	 -0.04941	& 3.2965$\times 10^{-4}$ & 3.4904$\times 10^{-4}$ & 1.1295$\times 10^{-4}$ \\
1,3 & -0.03242	& 1.3758$\times 10^{-6}$ & 1.8240$\times 10^{-5}$ & 1.5622$\times 10^{-5}$ \\
1,4 &	 -0.03153	& 7.8434$\times 10^{-8}$ & 1.1632$\times 10^{-5}$ & 5.3270$\times 10^{-6}$ \\
2,3 & -0.03885	& 5.7797$\times 10^{-5}$ & 1.0008$\times 10^{-4}$ & 7.6934$\times 10^{-5}$\\
2,4 & -0.03457  & 1.1048$\times 10^{-5}$ & 4.9776$\times 10^{-5}$ & 4.9492$\times 10^{-5}$ \\
3,4 & -0.03214  & 7.9189$\times 10^{-7}$ & 3.9694$\times 10^{-6}$ & 3.5507$\times 10^{-7}$ \\
1,2,3	 & -0.03824 & 4.8828$\times 10^{-5}$ & 1.1027$\times 10^{-4}$	& 9.2654$\times 10^{-5}$ \\
1,2,4	 & -0.03433 & 9.5012$\times 10^{-6}$ & 5.4966$\times 10^{-5}$	& 5.4882$\times 10^{-5}$ \\
1,3,4	 & -0.03196 & 4.9800$\times 10^{-7}$ & 8.7447$\times 10^{-6}$	& 4.3961$\times 10^{-6}$ \\
2,3,4	 & -0.03422 & 8.8319$\times 10^{-6}$ & 3.7932$\times 10^{-5}$	& 3.7899$\times 10^{-5}$ \\
1,2,3,4 & -0.03404 &	7.7898$\times 10^{-6}$ & 2.8065$\times 10^{-5}$ & 2.8065$\times 10^{-5}$ \\
\end{tabular}
\end{ruledtabular}
\end{table*}

\begin{figure*}
\caption{The exact (dotted lines) and approximate (solid lines) wave functions corresponding to the $4S$ state of the hydrogen atom are depicted on the 
left-hand side of the figure on different radial distance scales. The exact wave function has nodes at $r = 1.8716$, 6.6108, and 15.5180 bohr, while the approximate wave function has nodes at 2.0240, 6.6068, and 15.6442 bohr. The four nodal regions are enumerated from the nucleus outwards. The exact energy of the hydrogen atom in the $4S$ state is $-0.03125\ E_h$, while the nodal-region energies of the approximate hydrogen-atom wave function are $-14.010\times 10^{-2}\ E_h$ (first), $-1.000\times 10^{-2}\ E_h$ (second), $-3.261\times 10^{-2}\ E_h$ (third), and $-3.106\times 10^{-2}\ E_h$ (fourth). The corresponding local energies are depicted on the right-hand side of the figure [kinetic: green (upper) curves; potential: blue (bottom) curve; total: red (middle) lines].}
\includegraphics[scale=0.85]{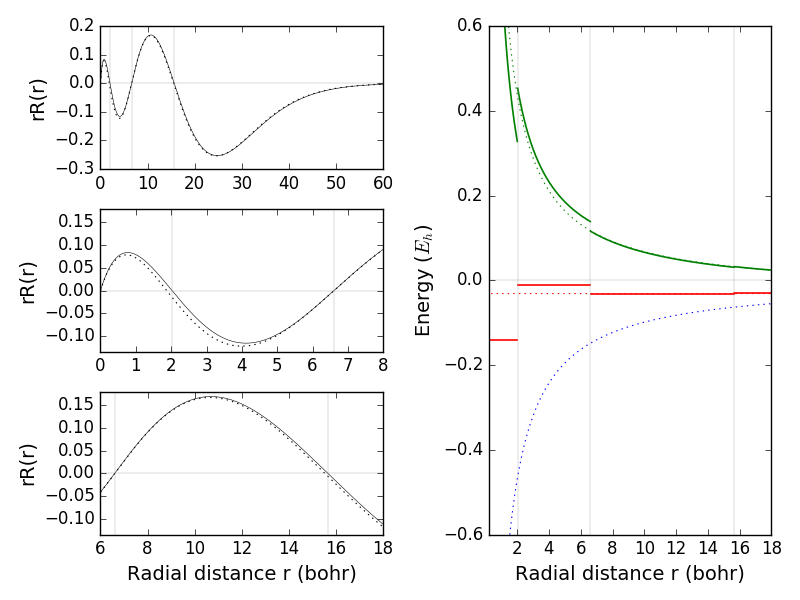}
\end{figure*}

\begin{figure*}
\caption{The exact (dotted lines) and approximate (solid lines) wave functions corresponding to the fifth excited state of the harmonic oscillator (HO) are depicted on the left-hand side of the figure. The exact wave function (upper and lower left) has nodes at 0.959 and 2.020, while the first approximate wave function (HO-1, upper left) has nodes at 0.759 and 2.080 and the second approximate wave function (HO-2, lower left) has nodes at 0.985 and 2.420. The exact energy of HO is 5.5000, while the nodal-region energies of the HO-1 are 8.6564, 3.8478, and 5.6742 and the nodal-region energies of HO-2 are 5.2218, 3.8525, and 6.7234. The corresponding local energies are depicted on the right-hand side of the figure. The kinetic components are in green, the potential components are in blue and the sum of the two is in red.  Only the right halves of the wave functions are shown due to the antisymmetry with respect to the origin. A representation of the harmonic-oscillator problem with unitless distance and energy is chosen [24].}
\includegraphics[scale=0.85]{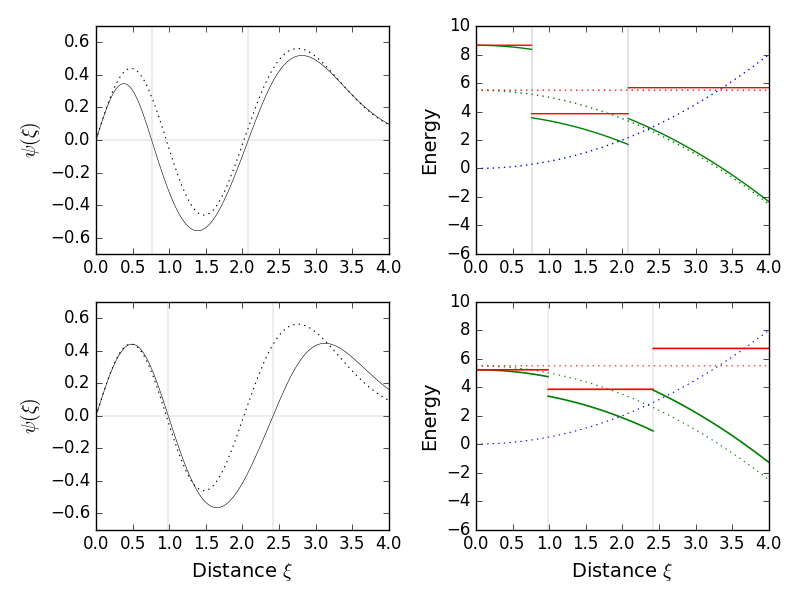}
\end{figure*}

It was observed earlier that a many-electron wave function,
with a domain of consideration that is restricted to a single
nodal region, is equivalent to a single-electron wave function
in a higher-dimensional space. As a result, a single-electron
example is worthwhile for demonstrating the qualitative features
of approximate nodal regions.As an illustration, consider
the exact and approximate 4$S$ state of the hydrogen atom. The
approximate wave functions minimize the total energy, while
being constrained to approximate nodes. Tables I-IV  present
the energies and corresponding errors of the minimizing wave
functions in single, double, triple, and quadruple combinations
of nodal regions for four 4$S$ state wave functions of the
hydrogen atom with approximate nodes. The utility of error
expression (\ref{10}) for helping to select the best wave function is
reflected in the fact that the wave function with the best average
energy, which is associated with the bottom row of Table III,
is the one that gives the lowest value for error expression (\ref{10});
compare the bottom rows of Tables I-IV. Comparison of the
bottom rows in Table I and Table IV also reveals, however, that
it is possible for a wave function with a higher value for error
expression (\ref{10}) to actually give a better average energy.

The red (middle) lines on the right-hand side of Fig. 1 [23],
which is associated with the wave function in Table I, depict the
nodal-region energy minima (solid lines) and actual eigenvalue
(dotted line) of the example. The ground-state energies in the
different approximate nodal regions are not necessarily equal,
making the energy discontinuous across the nodes.

The right-hand side of Fig. 1 depicts the split of the nodal-region energies into their local kinetic- and potential- energy components, obtained by rearranging the eigenvalue equation as $\frac{-\frac{1}{2}\nabla^2 \Psi(\vec{r})}{\Psi(\vec{r})}+V(\vec{r})=E$, which is the way it is utilized, for instance, in the familiar local energy and variance expressions [25, 26, 27]. As can be seen in Fig. 1, about the same nodal deviation from an exact nodal position can have a dramatically different impact on the nodal energy, depending on the strength of the external potential   at the position of a node. For this reason, energy-based error expressions (\ref{10}) and (\ref{12}) give measures for gauging nodal quality that should provide worthwhile alternatives to the use of the geometric notion of nodal distance error [28, 29]. These energy-based error expressions measure the cumulative deviation of the nodal-region energy minima from the average energy. When the nodes are exact, all of these nodal-region energy minima equal the excited-state eigenvalue, which is a constant throughout the entire space. 

It is interesting to note that the value of error expression (\ref{10}) can be determined solely by the discontinuities of the local kinetic energy at the nodes. The value of error expression (\ref{10}) is invariant with respect to a shift of all the nodal-region energy minima by the same constant and, as a result, this value depends only on the differences of the nodal-region energy minima. The differences of the neighboring nodal-region energy minima are, in turn, equivalent to the extent of the discontinuities of the local kinetic energy at the respective nodes.

\begin{table*}
\caption{The energies and error-expression evaluations of the approximate wavefunctions (HO-1 and HO-2 as defined previously) are shown in the table. The nodal regions, where the wavefunctions are considered, are indicated in the leftmost column (in parentheses). The harmonic oscillator energies are unitless.}
\begin{ruledtabular}
\begin{tabular}{cccc}
Approx. wave func. & Energy & Error expr. 1,& Error expr. 2,\\
&&Eq. (\ref{10}) &Eq. (\ref{12}), [29]\\
\hline
HO-1 (1,2,3) & 5.1974 & 1.9478 & 4.6713 \\
HO-1 (3) & 5.6742 & 0.2273 & 0.2273 \\
HO-2 (1,2,3) & 5.1319 & 1.6451 & 1.3926 \\
HO-2 (1) & 5.2218 & 0.0081 & 0.0081 \\
\end{tabular}
\end{ruledtabular}
\end{table*}

If, through the use of error expressions (\ref{10}) or (\ref{12}), there is an indication that a particular subset of nodal regions might be preferred, then it would be reasonable to consider choosing this subset alone. For example, for the wave function associated with Table I, if only the third and fourth nodal regions (as counted from the nucleus outwards) are used, instead of all four nodal regions, the values of the error expressions of Eq. (\ref{10}) and Eq. (\ref{12}) [30] go down from $1.0167\times10^{-4}\ E_h^2$ to $2.9484\times 10^{-7}\ E_h^2$ and from $3.0694\times 10^{-3}\ E_h^2$ to $7.9837\times 10^{-7}\ E_h^2$, respectively, where $E_h$ signifies the Hartree unit of energy. Simultaneously, the approximate energy estimate improves from $-0.03139\ E_h$ to $-0.03126\ E_h$ compared with the exact value of $-0.03125\ E_h$. It becomes clear that a restriction to the third and fourth nodal regions of the approximate hydrogen atom $4S$ wave function improves the energy estimate. In fact, compared with all the nodal combinations in Table I, the use of the third and fourth regions gives both the lowest value for error expression (\ref{10}) and the best average energy. 

In addition to the exact and approximate wave functions for
the 4S state of the hydrogen atom, consider also the exact and
two approximate fifth excited states of the one-dimensional
harmonic oscillator \footnote{The two wave functions that minimize the energy-expectation value
of the harmonic oscillator, while being constrained to nodes displaced
from the exact positions, are abbreviated HO-1 and HO-2 (see the
caption of Fig. 2 for more details).} (Fig. 2) [23].

Table V summarizes the energy and error expression values
of the minimizing wave functions with nodal approximations
both in the entire space and in selected nodal regions only.
It becomes clear that a restriction of HO-1 to the third nodal
region and of HO-2 to the first nodal region improves the energy
estimates.

While the mathematical results in this paper are general, the
difficulty is that their applications to many-electron systems
require flexible and robust numerical representations of the
multidimensional nodes. For these purposes, one might use
generalizations of the approach in Ref. [31]. In any case, our
theorem justifies the interpretation that approximate excited state
energies and wave functions are obtained even when the
exact nodes are only known approximately, as exemplified by
the cases given in Sec. I.

\section{CONCLUDING REMARKS}

In this paper, a minimum principle featuring nodes was
proven for excited states. Aspects of this minimum principle
are currently being actively utilized in practice, but here we
provide a proof.

The excited-state theorem within provides the realization
that the minimization over the entire space can be partitioned
into interconnected minimizations in individual exact nodal
regions, and an energy minimization over all space or over
one or several nodal regions gives the exact excited-state
energy. Moreover, the exact excited-state wave function is
obtained when the minimization is performed over all space.
The smoothness of the trial wave functions across the nodes
is the \emph{essential link} between the minimizations within 
each of the nodal regions \emph{for the construction of the correct 
minimizing wave function}, which is needed for the computation of
properties. Explicit expressions for the wave-function variation
around an excited state with the nodes constrained to the correct
ones are given in Eqs. (6) and (8).

Expressions (10) and (12) of the corollaries extend the minimum
principle to nodal variations when the exact nodes are
unknown. The lemma supports the corollaries and establishes
a key connection between the nodes and the antisymmetry of
the minimizing wave functions.

The main results in this paper are formulated in the theorem,
antisymmetry lemma, and corollaries. In addition, we have
provided suggestions for calculations of excited states when
approximate nodes are used in the nodal energy minimization
process. With this in mind, simple numerical examples illustrate
the use of expressions (10) and (12) as error estimates of
approximate nodes.

It is expected that the excited-state minimum principle
presented here and the extension of the minimum principle
to nodal variation will have a wide range of new applications
due to the general validity of these principles for excited states.\\

\begin{acknowledgments}
M.L. is grateful to William E. Palke for discussions years ago. 
This research was supported by the U.S. Department of
Energy, Office of Basic Energy Sciences, Division of Chemical
Sciences, Geosciences, and Biosciences through the Ames
Laboratory. The Ames Laboratory is operated for the U.S.
Department of Energy by Iowa StateUniversity under Contract
No. DE-AC02-07CH11358.
\end{acknowledgments}

\appendix

\section{}

The nodal constraint is imposed by restricting the variational space to the linear space of wave functions that are well behaved and have the nodes of 
the $k$th eigenfunction $\Psi_k(\textbf r_1,\textbf r_2,...,\textbf r_N)$. See Ref. [22] for a definition of an analytically
well behaved wave function. In the case of approximate
nodes, slightly weaker conditions are assumed, namely, that
the wave function is well behaved in the above sense in each
nodal region and only first-order smooth, i.e., the wave function
has continuous first derivatives, at the nodes. The restricted variational space is linear, as a linear combination of such trial wave functions is still a wave function with the properties that are assumed above. Alternative to restricting the variational space, the Hamiltonian of interest, $\hat{H}=\hat{T}+\hat{V}$ [the kinetic part is $\hat{T}=-\frac{1}{2}\sum_{1\le i\le N}\nabla_i^2$, where $\nabla_i=\frac{\partial }{\partial r_{i,x}}+\frac{\partial }{\partial r_{i,y}}+\frac{\partial }{\partial r_{i,z}}$ is acting on the i-the electronic coordinates, and the potential part is $\hat V=\sum_{1\le i<j\le N}\frac{1}{|\textbf r_i-\textbf r_j|}+\sum_{1\le j\le N}v(\textbf r_i) $, where $v(\textbf r)$ is the external potential], might be modified with the addition of $\delta$-function-type infinite potential walls along the nodes. A replacement of $\hat{H}$ with such a modified Hamiltonian $\hat{H}'$ is an alternative way to ensure a nodal constraint on the trial wave functions upon energy minimization, as the eigenfunctions of $\hat{H}'$ naturally have nodes at the places where the potential of $\hat{H}'$ becomes infinite.  

\section{}

Here are the details for the derivation of Eq. (\ref{7}).

\begin{widetext}
\begin{align}\tag{B1}\label{B1}
\langle g\Psi_k|\hat T|g\Psi_k\rangle-\langle g^2\Psi_k|\hat T|\Psi_k\rangle
&=-\frac{1}{2}\sum_{i=1}^N\sum_{\alpha=x,y,z}\langle g\Psi_k|\frac{\partial^2}{\partial r_{i,\alpha}^2}|g\Psi_k\rangle
+\frac{1}{2}\sum_{i=1}^N\sum_{\alpha=x,y,z}\langle g^2\Psi_k|\frac{\partial^2}{\partial r_{i,\alpha}^2}|\Psi_k\rangle\nonumber\\
&=\frac{1}{2}\sum_{i=1}^N\sum_{\alpha=x,y,z}
\langle \frac{\partial (g\Psi_k)}{\partial r_{i,\alpha}}|\frac{\partial (g\Psi_k)}{\partial r_{i,\alpha}}\rangle
-\frac{1}{2}\sum_{i=1}^N\sum_{\alpha=x,y,z}
\langle \frac{\partial (g^2\Psi_k)}{\partial r_{i,\alpha}}|\frac{\partial \Psi_k}{\partial r_{i,\alpha}}\rangle\nonumber\\
&=\frac{1}{2}\sum_{i=1}^N\sum_{\alpha=x,y,z}
\langle (\frac{\partial g}{\partial r_{i,\alpha}})\Psi_k+g(\frac{\partial\Psi_k}{\partial r_{i,\alpha}})|
(\frac{\partial g}{\partial r_{i,\alpha}})\Psi_k+g(\frac{\partial\Psi_k}{\partial r_{i,\alpha}})
\rangle\nonumber\\
&-\frac{1}{2}\sum_{i=1}^N\sum_{\alpha=x,y,z}
\langle (2g\frac{\partial g}{\partial r_{i,\alpha}})\Psi_k+g^2(\frac{\partial\Psi_k}{\partial r_{i,\alpha}})|
\frac{\partial\Psi_k}{\partial r_{i,\alpha}}\rangle\nonumber\\
&=\frac{1}{2}\sum_{i=1}^N\sum_{\alpha=x,y,z}\langle (\frac{\partial g}{\partial r_{i,\alpha}})\Psi_k|
(\frac{\partial g}{\partial r_{i,\alpha}})\Psi_k\rangle\nonumber\\
&=\frac{N_{\uparrow}}{2}\sum_{\alpha=x,y,z}\langle (\frac{\partial g}{\partial r_{N_{\uparrow},\alpha}})\Psi_k|
(\frac{\partial g}{\partial r_{N_{\uparrow},\alpha}})\Psi_k\rangle\nonumber\\
&+\frac{N_{\downarrow}}{2}\sum_{\alpha=x,y,z}\langle (\frac{\partial g}{\partial r_{N_{\uparrow}
+N_{\downarrow},\alpha}})\Psi_k|(\frac{\partial g}{\partial r_{N_{\uparrow}+N_{\downarrow},\alpha}})\Psi_k\rangle\nonumber.
\end{align}
\end{widetext}

The following arguments are used in Eq. (\ref{B1}): (1) integration by parts
in the second equality, (2) derivative of a product in the third
equality, (3) algebraic simplification in the fourth equality, and (4)
coordinate interchange symmetry of $g(\textbf r_1,\textbf r_2,...,\textbf r_N)$, if there is such symmetry, in the last equality.

\section{}

This appendix demonstrates that
$g(\textbf r_1,\textbf r_2,...,\textbf r_N)=\frac{\Psi^{(k)}(\textbf r_1,\textbf r_2,...,\textbf r_N)}
{\Psi_k(\textbf r_1,\textbf r_2,...,\textbf r_N)}$
 is finite,
assuming both the eigenfunction
$\Psi_k(\textbf r_1,\textbf r_2,...,\textbf r_N)$ and the trial
wave function $\Psi^{(k)}(\textbf r_1,\textbf r_2,...,\textbf r_N)$ 
are analytic around the node.

An eigenfunction has 3$N$ variables and its node, i.e. the positions in
the 3$N$-dimensional space where the wave function is zero, is a
hypersurface of dimension (3$N$-1). For each point on the nodal
hypersurface there is a one-dimensional direction, perpendicular to the
nodal hypersurface, that leads toward non-zero values, so the behavior
of the eigenfunction, in the vicinity of its node, is effectively
described by a one-dimensional Schrodinger equation:

\begin{align}\tag{C1}\label{C1}
\frac{d^2\Psi_k(r)}{dr^2}=f(r)\Psi_k(r),
\end{align}

\noindent where $f(r)=-2[ E_k-V(r)]$.
Subsequent differentiation of Eq. (\ref{C1}) gives

\begin{widetext}
\begin{align}\tag{C2}\label{C2}
\frac{d^3\Psi_k(r)}{dr^3}&=\frac{df(r)}{dr}\Psi_k(r)+f(r)\frac{d\Psi_k(r)}{dr}\\
\frac{d^4\Psi_k(r)}{dr^4}&=\frac{d^2f(r)}{dr^2}\Psi_k(r)+2\frac{df(r)}{dr}\frac{d\Psi_k(r)}{dr}+f(r)\frac{d^2\Psi_k(r)}{dr^2}\nonumber\\
&\dots\nonumber\ .
\end{align}
\end{widetext}

Now, we employ a proof by contradiction. If
$\frac{d\Psi_k(r)}{dr}|_{r=0}=0$ as well as
$\Psi_k(0)=0$, then Eqs.
(\ref{C1}) and (\ref{C2}) dictate that all higher derivatives of the
eigenfunction also vanish, i.e.
$\frac{d^n\Psi_k(r)}{dr^n}|_{r=0}=0$ for any $n$.
Based on the assumption that
$\Psi_k(r)$ is analytic
around the node at
$r=0$, it follows
that the eigenfunction identically vanishes everywhere around the
origin, i.e. $\Psi_k(r)\equiv 0$,
which is absurd. Consequently,
$\Psi_k(0)=0$ but
$\frac{d\Psi(0)}{dr}\not=0$. Hence,
assuming the eigenfunction can be expanded in a Taylor series around the
point at the node
($r=0$),
$\Psi_k(r)=a_1r+a_2r^2+a_3r^3+...=r(a_1+a_2r+a_3r^2+...)$, where
$a_1\not=0$.

The Taylor expansion of a trial wave function around a point at the node
has to be $\Psi^{(k)}(r)=b_nr^n+b_{n+1}r^{n+1}+b_{n+2}r^{n+2}+...=r^n(b_n+b_{n+1}r+b_{n+2}r^2+...)$,
where $b_n\not=0$ and
$n\ge 1$. The
prefactor $r^n$
guarantees the trial wave function
$\Psi^{(k)}(r)$ vanishes at
the node ($r=0$).

As a result, $\frac{\Psi^{(k)}(r)}{\Psi_k(r)}=\frac{r^n(b_n+b_{n+1}r+b_{n+2}r^2+...)}{r(a_1+a_2r+a_3r^2+...)}
=\frac{r^{n-1}(b_n+b_{n+1}r+b_{n+2}r^2+...)}{a_1+a_2r+a_3r^2+...}$
does not diverge at the node of the eigenfunction.

\bibliography{apssamp}

\begin{quote}
\textbf{References:}
\end{quote}

1. 1.	E. Hylleraas and B. Undheim, Z. Phys. \textbf{65}, 759 (1930); 
J. K. L. McDonald, Phys. Rev. \textbf{43}, 830 (1933). 

2. R. Courant and D. Hilbert, \emph{Methods of Mathematical Physics},
1\textsuperscript{st} English ed. (Interscience, New York, 1953).

3. J. O. Hirshfeld, C. J. Goebel, and L. W. Bruch, J. Chem. Phys.
\textbf{61}, 5456 (1974); E. B. Wilson, \emph{ibid.} \textbf{63}, 4870
(1975); G. Hunter, Int. J. Quantum Chem. \textbf{19}, 755 (1981).

4. P. J. Reynolds, D. Ceperley, B. J. Alder, and W. A. Lester, Jr., J.
Chem. Phys. \textbf{77}, 5593 (1982).

5. D. Ceperley, J. Stat. Phys. \textbf{63}, 1237 (1991).

6. W. M. C. Foulkes, R. Q. Hood, and R. J. Needs, Phys. Rev. B
\textbf{60}, 4558 (1999).

7. P. G. Hipes, Phys. Rev. B \textbf{83}, 195118 (2011).

8. M. Bajdich, L. Mitas, G. Drobny, and L. K. Wagner, Phys. Rev. B
\textbf{72}, 075131 (2005).

9. L. Mitas, Phys. Rev. Lett. \textbf{96}, 240402 (2006).

10. J. S. Briggs and M. Walter, Phys. Rev. A \textbf{74}, 062108 (2006).

11. D. Bressanini and P. J. Reynolds, Phys. Rev. E \textbf{84}, 046705
(2011).

12. A. J. Williamson, J. C. Grossman, R. Q. Hood, A. Puzder, and G.
Galli, Phys. Rev. Lett. \textbf{89}, 196803 (2002); J. E. Vincent, J. Kim, and R.
M. Martin, Phys. Rev. B \textbf{75}, 045302 (2007).

13. J. Yu, L. K. Wagner, and E. Ertekin, J. Chem. Phys. \textbf{143}, 224707
(2015).

14. R. Q. Hood, P. R. C Kent, R. J. Needs, and P. R. Briddon, Phys. Rev.
Lett. \textbf{91}, 076403 (2003).

15. F. Marsusi, J. Sabbaghzadeh, and N. D. Drummond, Phys. Rev. B \textbf{84},
245315 (2011).

16. A. Ghosla, A. D. Guclu, C. J. Umrigar, D. Ullmo, and H. U. Baranger,
Phys. Rev. B \textbf{76}, 085341 (2007); S. A. Blundell and S. Chacko, \emph{ibid.} \textbf{83}, 195444 (2011).

17. N. D. Drummond and R. J. Needs, Phys. Rev. B \textbf{87}, 045131 (2013).

18. T. Bouabca, N. B. Amor, D. Maynau, and M. Caffarel, J. Chem. Phys.
\textbf{130}, 114107 (2009); N. Dupuy, S. Bouaouli, F. Mauri, S. Sorella, and M.
Casula, \emph{ibid.} \textbf{142}, 214109 (2015); R. Guareschi,
H. Zulfikri, C. Daday, F. M. Floris, C. Amovilli, B. Mennucci,
and C. Filippi, J. Chem. Theory Comput. \textbf{12}, 1674 (2016).

19. A. S. Petit, B. A. Wellen, and A. B. McCoy, J. Chem. Phys \textbf{138},
034105 (2013); R. C. Fortenberry, Q. Yu, J. S. Mancini, J. M. Bowman, T.
J. Lee, T. Daniel Crawford, W. F. Klemperer, and J. S. Francisco, \emph{ibid.} \textbf{143}, 071102 (2015); J. E. Ford and A. B. McCoy, Chem. Phys.
Lett. 645, \textbf{15} (2016).

20. C.-J. Huang, C. Filippi, and C. Umrigar, J. Chem. Phys.
\textbf{108}, 8838 (1998).

21. R. McWeeny, \emph{Methods of Molecular Quantum Mechanics}, (Academic, New York, 1989).

22. R. D. Richtmyer, \emph{Principles of Advanced Mathematical Physics},
Vol. I (Springer-Verlag, Berlin - 1978).

23. The numerical calculations were done using modified versions of harmonic1.f90 and hydrogen$\_$radial.f90 from [24].

24. P. Giannozzi, Lecture notes in  numerical methods in quantum mechanics, 
http://www.fisica.uniud.it/?giannozz/Corsi/MQ/mq.html (unpublished).

25. C. J. Umrigar, K. G. Wilson, and J. W. Wilkins, Phys. Rev. Lett. \textbf{60}, 1719 (1988). 

26. P. R. C. Kent, R. J. Needs, and G. Rajagopal, Phys. Rev. B \textbf{59,} 12344 (1999). 

27. J. H. Bartlett, Phys. Rev. \textbf{98}, 1067 (1955).   

28. F. A. Reboredo and P. R. C. Kent, Phys. Rev. B \textbf{77}, 245110 (2008). 

29. M. Dubecky, R. Derian, L. Mitas, and I. Stich, J. Chem. Phys. \textbf{133}, 244301 (2010). 

30. The expression $\frac{1}{M}\sum_{n=1}^M\Bigg[\frac{\langle g_n\tilde{\Psi}_{min}^{(k)}|\hat H|\tilde{\Psi}_{min}^{(k)}\rangle}{\langle g_n\tilde{\Psi}_{min}^{(k)}|\tilde{\Psi}_{min}^{(k)}\rangle}
-\langle\tilde{\Psi}_{min}^{(k)}|\hat H|\tilde{\Psi}_{min}^{(k)}\rangle\Bigg]^2$ is used as a substitute for expression (\ref{12}), which employs all possible scaling functions. The $M$ scaling functions are of the 
form $g=e^{-2d(x-x_0)^2}$. The parameters for the hydrogen atom example are ($M=4$) as follows: $d_1=12.3251$, $x_{0,1}=0.9358$; $d_2=1.9222$, $x_{0,2}=4.2412$; $d_3=0.3527$, $x_{0,3}=11.0644$; and $d_4=0.0104$, $x_{0,4}=47.7590$, while the ones for the two harmonic oscillator examples are ($M=3$): $d_1=46.9440$, $x_{0,1}=0.4795$; $d_2=38.3518$,  $x_{0,2}=1.4895$; and $d_3=1.7408$, $x_{0,3}=4.510$.

31. T. C. Scott, A. Luchow, D. Bressanini, and J. D. Morgan III, Phys. Rev. A \textbf{75}, 060101(R) (2007).

\end{document}